\documentclass[runningheads]{llncs}
\usepackage{graphicx}
\usepackage[misc,geometry]{ifsym} 

\makeatletter
\def\thickhline{%
  \noalign{\ifnum0=`}\fi\hrule \@height \thickarrayrulewidth \futurelet
   \reserved@a\@xthickhline}
\def\@xthickhline{\ifx\reserved@a\thickhline
               \vskip\doublerulesep
               \vskip-\thickarrayrulewidth
             \fi
      \ifnum0=`{\fi}}
\makeatother

\newlength{\thickarrayrulewidth}
\begin{document}

\newcommand\mypound{\scalebox{0.8}{\raisebox{0.4ex}{\#}}}
\newcommand{\bb}[1]{\mathbb{#1}}

\title{Don't Forget to Lock the Front Door! \\ Inferring the Deployment of Source Address Validation of Inbound Traffic}

\author{Maciej Korczy\'nski\inst{1}\textsuperscript{(\Letter)} 
\and
Yevheniya Nosyk\inst{1} \and
Qasim Lone \inst{2} \and Marcin Skwarek\inst{1} \and Baptiste Jonglez\inst{1} \and Andrzej Duda\inst{1}}

\authorrunning{ }

\institute{Univ. Grenoble Alpes, CNRS, Grenoble INP, LIG, F-38000 Grenoble, France\\
\email{maciej.korczynski@univ-grenoble-alpes.fr}\\
\and
Delft University of Technology, The Netherlands}

\titlerunning{ }
\pagenumbering{gobble}

\maketitle            

\begin{abstract}
This paper concerns the problem of the absence of ingress filtering at the network edge, 
one of the main causes of important network security issues. Numerous network operators do not deploy the best current practice---Source Address Validation (SAV) that aims at mitigating these issues. 
We perform the first Internet-wide active measurement study to enumerate networks not filtering \textit{incoming packets} by their source address. The measurement method consists of identifying closed and open DNS resolvers handling requests coming from the outside of the network with the source address from the range assigned inside the network under the test. The proposed method provides the most complete picture of the \textit{inbound} SAV deployment state at network providers. We reveal that 32~673 Autonomous Systems (ASes) and 197~641 Border Gateway Protocol (BGP) prefixes are vulnerable to spoofing of inbound traffic. Finally, using the data from the Spoofer project and performing an open resolver scan, we compare the filtering policies in both directions. 

\end{abstract}

\section{Introduction}
The Internet relies on IP packets to enable communication between hosts with the destination and source addresses specified in packet headers. However, there is no packet-level authentication mechanism to ensure that the source addresses have not been altered~\cite{Beverly:2009:UED:1644893.1644936}. The modification of a source IP address is referred to as ``IP spoofing''. It results in the anonymity of the sender and prevents a packet from being traced to
its origin. This vulnerability has been leveraged to launch Distributed Denial-of-Service (DDoS) attacks, in particular, the reflection attacks~\cite{bb-spoofer-sruti}. Due to the absence of a method to block packet header modification, the efforts have been undertaken to prevent spoofed packets from reaching potential victims. This goal can be achieved with packet filtering at the network edge, formalized in RFC 2827 and called \textit{Source Address Validation} (SAV)~\cite{Ferguson:2000:NIF:RFC2827}. 

The role of IP spoofing in cyberattacks drives the need to estimate the level of SAV deployment by network providers. There exist projects aimed at enumerating networks without packet filtering, for example, the Spoofer~\cite{Spoofer}. However, a great majority of the existing work concentrates on \textit{outbound} SAV, the root of DDoS attacks~\cite{Kuhrer:2014:EHR:2671225.2671233}. While less obvious, the lack of \textit{inbound} filtering enables an attacker to appear as an internal host of a network and may reveal valuable information about the network infrastructure, not seen from the outside. Inbound IP spoofing may serve as a vector for cache poisoning attacks~\cite{kaminsky} even~if~the 
Domain Name System (DNS) server is correctly configured as a closed~resolver.

In this work, we report on the preliminary results of the Closed Resolver Project \cite{closed}. We propose a new method to identify networks not filtering inbound traffic by source IP addresses. We perform an Internet-wide scan of BGP prefixes maintained by RouteViews~\cite{routeviews} for the IPv4 address space to identify closed and open DNS resolvers in each routable network. We achieve this goal by sending DNS A requests with spoofed source IP addresses for which the destination is every 
host of every routing prefix and the source is the next host from the same network. If there is no filtering in transit networks and at the network edge, such a packet is received by a server that resolves the DNS A request for a host that seems to be on the same network. As our scanner spoofs the source IP address, the response from the local resolver is directly sent to the spoofed client IP address, preventing us from analyzing 
it. However, we control the authoritative name server for the queried domains and observe from which networks it receives the requests. This method identifies networks not performing filtering of \textit{incoming packets} without the need for a vantage point inside the network itself.

The above method when applied alone shows the absence of inbound SAV at the network edge. In parallel, we send subsequent unspoofed DNS A record requests to identify open resolvers at the scale of the Internet. 
If open resolvers reply to the unspoofed requests but not to the spoofed ones, we infer the presence of SAV for incoming traffic
either at the network edge or in transit networks. 
By doing this, we detect both the absence and the presence of inbound packet~filtering. 

We analyze the geographical distribution of 
networks vulnerable to inbound spoofing and identify the countries that do not comply with the SAV standard, which is the first step in mitigating the issue by contacting 
Computer Security Incident Response Teams (CSIRTs).

We also retrieve the Spoofer data and deploy a method proposed by Mauch~\cite{mauch} to infer the absence and the presence of \textit{outbound} SAV. In this way, we study the policies of the SAV deployment per provider in both directions. Previous work demonstrated the difficulty in incentivizing providers to deploy filtering for outbound traffic due to misaligned economic incentives: implementing SAV for outbound traffic benefits other networks and not the network of the deployment \cite{spoofer_new}. This work shows how the deployment of SAV for inbound traffic protects the provider's own network.

\vspace{-0.2cm}
\section{Background}
Source address validation was proposed in 2000 in RFC 2827 as a result of a growing number of DDoS attacks. The RFC defined the notion of ingress filtering---discarding any packets with source addresses not following filtering rules. This operation is the most effective when applied at the network edge~\cite{Ferguson:2000:NIF:RFC2827}. RFC 3704 proposed different ways to implement SAV including static access control lists (ACLs) and reverse path forwarding \cite{Baker:2004:IFM:RFC3704}. Packet filtering can be applied in two directions: inbound to a customer (coming from the outside to the customer network) and outbound from a customer (coming from inside the customer network to the outside). 
The lack of SAV in any of these directions may result in different~security~threats.

Attackers benefit from the absence of outbound SAV to launch DDoS attacks, in particular, amplification attacks. Adversaries make use of public services prone to amplification~\cite{hell} to which they send requests on behalf of their victims by spoofing their source IP addresses. The victim is then overloaded with the traffic coming from the services rather than from the attacker. In this scenario, the origin of the attack is not traceable. One of the most successful attacks against GitHub resulted in traffic of 1.35 Tbps: attackers redirected Memcached responses by spoofing their source addresses~\cite{github}. In such scenarios, spoofed source addresses are usually random globally routable IPs. In some cases, to impersonate an internal host, a spoofed IP address may be from the inside target network, which reveals the absence of inbound SAV~\cite{Baker:2004:IFM:RFC3704}.

Pretending to be an internal host reveals information about the inner network structure, such as the presence of closed DNS resolvers that resolve only on behalf of clients within the same network. Attackers can further exploit closed resolvers, for instance, for leveraging misconfigurations of the Sender Policy Framework (SPF)~\cite{email_prevention}. In case of not correctly deployed SPF, attackers can trigger closed DNS resolvers to perform an unlimited number of requests thus introducing a potential DoS attack vector. The possibility of impersonating another host on the victim network can also assist in the zone poisoning attack~\cite{Korczynski:2016:ZPN:2987443.2987477}. A master DNS server, authoritative for a given domain, may be configured to accept DNS dynamic updates from 
a DHCP server on the same network \cite{rfc2136}. Thus, sending a spoofed update from the outside with an IP address of that DHCP server will modify the content of the zone file \cite{Korczynski:2016:ZPN:2987443.2987477}. The attack may lead to domain hijacking. Another way to target closed resolvers is to perform DNS cache poisoning~\cite{kaminsky}. An attacker can send a spoofed DNS A request for a specific domain to a closed resolver, followed by forged replies before the arrival of the response from the genuine authoritative server. In this case, the users who query the same domain will be redirected to where the attacker specified until the forged DNS entry reaches~its Time To Live~(TTL).

Despite the knowledge of the above-mentioned attack scenarios and the costs of the damage they may incur, it has been shown that SAV is not yet widely deployed. Lichtblau et al. surveyed 84 network operators to learn whether they deploy SAV and what challenges they face~\cite{Lichtblau}. The reasons for not performing packet filtering include incidentally filtering out legitimate traffic, equipment limitations, and lack of a direct economic benefit. The last aspect assumes outbound SAV when the deployed network can become an attack source but cannot be attacked itself. Performing inbound SAV protects networks from, for example, the threats described above, which is beneficial from the economic perspective.

\begin{table}[t]
\caption{Methods to infer  deployment of Source Address Validation.} 
\label{related_work} 
\scriptsize
\centering 
\setlength{\tabcolsep}{7pt}
\setlength{\arrayrulewidth}{0.6pt}
\begin{tabular}{lcccc}
 \thickhline \\ [-1.5ex]
 \vspace{1mm}
\textbf{Method}&\textbf{Direction}&\textbf{Presence/}&\textbf{Remote}&\textbf{Relies on }\\[-1ex]
&&\textbf{Absence}&&\textbf{misconfigurations}\\[1ex]
\hline \\ [-1.5ex]
 Spoofer~\cite{bb-spoofer-sruti,Spoofer} & both & both & no & no\\
Forwarders-based~\cite{mauch,Kuhrer:2014:EHR:2671225.2671233} & outbound & absence & yes & yes\\
 Traceroute loops~\cite{loops} & outbound & absence & yes & yes\\
 Passive detection~\cite{Lichtblau,Lucas} & outbound & both & no & no\\
 Our method \cite{closed}& inbound & both & yes & no\\
 \noalign{\vskip 1ex}
 \thickhline
\end{tabular}
\end{table}

\section{Related Work \label{sec:related}}

Table~\ref{related_work} summarizes several methods proposed to infer SAV deployment. They differ in terms of the filtering direction (inbound/outbound), whether they infer the presence or absence of SAV, whether measurements can be done remotely or on a vantage point inside the tested network, and if the method relies on existing network misconfigurations.

The Spoofer project deploys a client-server infrastructure mainly based on volunteers (and ``crowdworkers'' hired for one study trough five crowdsourcing platforms~\cite{marketplaces}) that run the client software from inside a network. The active probing client sends both unspoofed and spoofed packets to the Spoofer server either periodically or when it detects a new network. The server inspects received packets (if any) and analyzes whether spoofing is allowed and to what extent~\cite{Beverly:2009:UED:1644893.1644936}. For every client running the software, its /24 IPv4 address block and the autonomous system number (ASN) are identified and measurement results are made publicly available\footnote{https://spoofer.caida.org/summary.php}. This approach identifies both the absence and the presence of SAV in both directions. The results obtained by the Spoofer project provide the most confident picture of the deployment of outbound SAV and have covered tests from 7~353 
ASes since 2015.
However, those that are not aware of this issue or do not deploy SAV are less likely to run Spoofer on their networks.

A more practical approach is to perform such measurements remotely. Kührer et al.~\cite{Kuhrer:2014:EHR:2671225.2671233} scanned for open DNS resolvers, as proposed by Mauch~\cite{mauch}, to detect the absence of outbound SAV. They leveraged the misconfiguration of forwarding resolvers. The misbehaving resolver forwards a request to a recursive resolver with either not changing the packet source address to its address or by sending back the response to the client with the source IP of the recursive resolver. They fingerprinted those forwarders and found out that they are mostly embedded devices and routers. Misconfigured forwarders originated from 2~692 autonomous systems. We refer to this technique as \textit{forwarders-based}.

Lone et al.~\cite{loops} proposed another method that does not require a vantage point inside a tested network. When packets are sent to a customer network with an address that is routable but not allocated, this packet is sent back to the provider router without changing its source IP address. The packet, having the source IP address of the machine that sent it, should be dropped by the router because the source IP does not belong to the customer network. The method detected 703 autonomous systems not deploying outbound SAV.

While the above-mentioned methods rely on actively generated (whether spoofed or not) packets, Lichtblau et al.~\cite{Lichtblau} passively observed and analyzed inter-domain traffic exchanged between more than 700 networks at a large IXP. They classified observed traffic into bogon, unrouted, invalid, and valid based on the source IP addresses and AS paths. The most conservative estimation identified 393 networks where the invalid traffic originated from.

We are the first to propose a method to detect the absence of inbound SAV that is remote and does not rely on existing misconfigurations. 
Instead, we use local DNS resolvers (both open and closed) to infer the absence of packet filtering and the presence of SAV either at transit networks or at the edge. 
We are aware of a similar method by Deccio, but his work is in progress and not yet publicly available~\cite{Casey}.

\begin{figure}[t]
\centering
\includegraphics[width=0.9\textwidth]{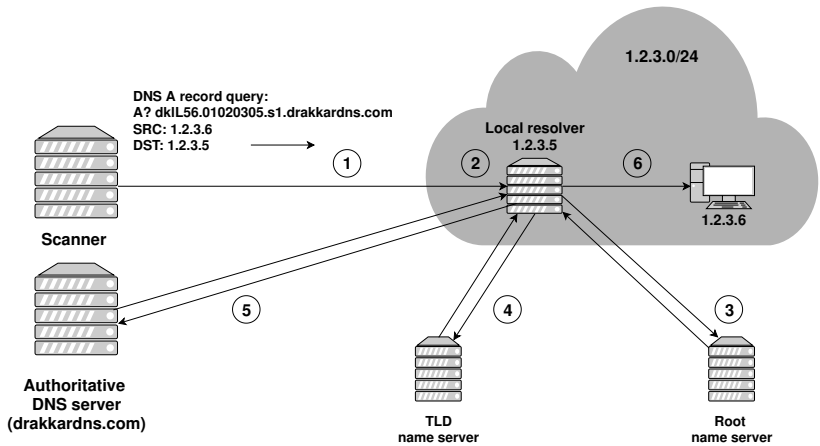}
\caption{Inbound spoofing scan setup.}
\label{setup}
\end{figure}

\section{Methodology}

\subsection{Spoofing Scan}

Figure~\ref{setup} illustrates the idea of the proposed method. We have developed an efficient scanner that sends hand-crafted DNS A record request packets. We run the scanner on a machine inside a network that does not deploy outbound SAV so that we can send packets with spoofed IP addresses\footnote{After our initial scan, we learned that one of the three upstream providers deploys SAV, so we temporarily disabled it to perform our measurements.}. 
We set up a DNS server authoritative for the \texttt{drakkardns.com} domain to capture the traffic related to our scans. 
When a resolver inside a network vulnerable to inbound spoofing performs query resolution, we observe it on our authoritative DNS server. To prevent caching and to be able to identify the true originator in case of forwarding, we query the following unique subdomain every time: a random string, the hex-encoded resolver IP address (the destination of our query), a scan identifier, and the domain name itself, for example, \texttt{qGPDBe.02ae52c7.s1.drakkardns.com}.

Figure~\ref{setup} shows the scanning setup for the \texttt{1.2.3.0/24} network. In step 1, the scanner sends one spoofed packet to each host of this network, thus packets to 254 destinations in total. The spoofed source IP address is always the next one after the destination. When the spoofed DNS packet arrives at the 
destination network edge (therefore it has not been filtered anywhere in transit), there are three possible cases:

\begin{list}{$\bullet$}{}  
    \item \textbf{Packet filtering in place}. The packet filter inspects the packet source address and detects that such a packet cannot arrive from the outside because the address block is allocated inside the network. Thus, the filter drops the packet.
    \item \textbf{No packet filtering in place and nothing prevents the packet from entering the network}. If the packet destination is \texttt{1.2.3.5}, the address of the local resolver (step 2), it receives a DNS A record request from what looks to be another host on the same network and performs query resolution. If the destination is not the local resolver, it will drop the packet. However, the scanner will eventually reach all the hosts on the network and the local resolver if there is one. In some cases, the closed DNS resolver may be configured to refuse queries coming from its local area network (for example, if the whole separate network is dedicated to the infrastructure).
    \item \textbf{Other cases}. Regardless of the presence or absence of filtering, packets may be dropped due to reasons not related to IP spoofing such as network congestion~\cite{Beverly:2009:UED:1644893.1644936}.
\end{list}

In this study, we distinguish between two types of local resolvers: forwarders (or proxies) that forward queries to other recursive resolvers and non-forwarders (non-proxies) that recursively resolve queries they receive. Therefore, the non-forwarding local resolver (\texttt{1.2.3.5}) inspects the query that looks as if it was sent from \texttt{1.2.3.6} and performs the resolution by iteratively querying the root (step 3) and the top-level domain name (step 4) servers until it reaches our authoritative DNS server in step 5.
Alternatively, it forwards the query to another recursive resolver that repeats the same procedure as described above for non-forwarders. In step 6, the DNS A query response is sent to the spoofed source (\texttt{1.2.3.6}).

We aim at scanning the whole IPv4 address space, yet taking into account only globally routable and allocated address ranges. We use the data maintained by the RouteViews Project to get all the IP blocks currently present in the BGP routing table and send spoofed DNS A requests to all the hosts of the prefixes.

\vspace{-0.1cm}
\subsection{Open Resolver Scan \label{sec:orscan}}

In parallel, we perform an open resolver scan by sending DNS A requests with the genuine source IP address of the scanner. 
To avoid temporal changes, we send a non-spoofed query just after the spoofed one to the same host. The format of a non-spoofed query is almost identical to the spoofed one. The only difference is the scan identifier: \texttt{qGPDBe.02ae52c7.n1.drakkardns.com}. If we receive a request on our authoritative DNS server, it means that we have reached an open resolver. Moreover, if this open resolver did not resolve a spoofed query, we infer the presence of inbound SAV either in transit or at the tested network edge.

We also analyze traffic on the machine on which we run the scanner to deploy the forwarders-based method, as explained in Section \ref{sec:related}. We distinguish between two cases: the source of the DNS response is the same as the original destination and the source is different~\cite{mauch,Kuhrer:2014:EHR:2671225.2671233}. The latter implies that either the source IP address of the original query was not rewritten when the query was forwarded to another recursive resolver or the source IP address of the recursive resolver was not changed on the way back. In either case, such a packet should not be able to leave its network if there is the outbound SAV in place.
In Section \ref{sec:policies}, we analyze the results from the forwarders-based method and compare the policies of SAV deployment in both directions.
\subsection{Ethical Considerations}
To make sure that our study follows the ethical rules of network scanning, yet providing complete results, we adopt the recommended best practices~\cite{eth,zmap}. We aggregate the BGP routing table to eliminate overlapping prefixes. In this way, we are sure to send no more than two DNS A request packets (spoofed and non-spoofed) to every tested host. 
Due to packet losses, we potentially miss some results, but we go with this limitation in order not to disrupt the normal operation of tested networks. In addition, we randomize our input list for the scanner so that we do not send consecutive requests to the same network (apart from two consecutive spoofed and non-spoofed packets). Our scanning activities are spread over 10 days.

We set up a homepage for this study on \texttt{drakkardns.com} and all queried subdomains with a description of our scan and provide the contact information if someone wants to exclude her networks from testing. 
We received 9 requests from operators and excluded 29~360~925 IP addresses from our future scans.
We also exclude those addresses from our analysis.
We do not publicly reveal the source address validation policies of individual networks and AS operators. We also plan a notification campaign through CSIRTs and by directly informing the operators of affected networks.

\vspace{-0.2cm}
\section{Results}

\subsection{Filtering Levels}

When evaluating the SAV deployment, we aim at finding the unit of analysis that will show the most consistent results. Each received request reveals the IP address of the original target of the query. We map this IP address to the corresponding /24 address block, the longest matching BGP prefix, and its ASN. 
This granularity allows us to evaluate the SAV practices at different~levels:

\begin{list}{$\bullet$}{}
    \item Autonomous systems: while based on a few received queries, we cannot by any means conclude on the filtering policies of the whole AS---they reveal SAV compliance for a part of it~\cite{bb-spoofer-sruti,Spoofer,loops,spoofer_new}.
    \item Longest matching BGP prefixes: as the provider ASes may sub-allocate their address space to their customers by prefix delegation~\cite{Krenc}, the analysis of the SAV deployment at the longest matching prefix is another commonly used unit of analysis~\cite{bb-spoofer-sruti,spoofer_new}. 
    \item /24 IPv4 networks: it is the smallest unit of measuring the SAV deployment used so far by the existing methods~\cite{Spoofer,spoofer_new}. We later show that even at this level some results are inconsistent.
\end{list}

\begin{table}[t]
\caption{Spoofing scan results} 
\label{global_results} 
\scriptsize
\centering 
\setlength{\tabcolsep}{8pt}
\setlength{\arrayrulewidth}{0.5pt}
\begin{tabular}{l|c|c}
 \thickhline 
\textbf{Metric} & \textbf{Number} & \textbf{Proportion (\%)}\\
 \thickhline
DNS forwarders & 6~530~799 & 94.01 \\
\hspace{8mm}Open resolvers & 2~317~607 & 35.49 \\
\hspace{8mm}Closed resolvers & 4~213~192 & 64.51 \\
\hline
DNS non-forwarders & 415~983 & 5.99 \\
\hspace{8mm}Open resolvers & 39~924  & 9.6 \\
\hspace{8mm}Closed resolvers & 376~059 & 90.4 \\

\hline 
Vulnerable to spoofing /24 IPv4 networks & 959~666 & 8.62 \\ 
Vulnerable to spoofing longest matching prefixes & 197~641 & 23.61 \\ 
Vulnerable to spoofing autonomous systems & 32~673 & 49.34 \\
 \thickhline
\end{tabular}
\end{table}

\subsection{Global Scans}

In December 2019, we performed the spoofing and open resolver scans. During the spoofing scan, we observed 14~761~484 A requests on our authoritative DNS server. It has been shown that DNS resolvers tend to issue repetitive queries due to proactive caching or premature querying~\cite{revealed}. Thus, we leave only unique tuples of the source IP address and the domain name, which results in 9~558~190 unique requests (64.75\% of all the received requests and 0.34\% of all the requests sent by the scanner).

Table~\ref{global_results} presents the statistics gathered from the spoofing scan. From each request received on our authoritative name server, we retrieve the queried domain, extract its hexadecimal part (the destination of our original DNS A query) and convert it to an IP address. We then compare it to the source IP of the query and identify 6~530~799 DNS proxies (local resolvers that forwarded their queries to other recursive resolvers) and 415~983 non-proxies (local resolvers that performed resolutions themselves). We identify that 35.49\% of forwarders and 9.60\% of non-forwarders are open resolvers. 

The address encoded in the domain name identifies the originator network. We map every IP address to the corresponding prefix and the autonomous system number. They originate from 32~673  autonomous systems and correspond to 197~641 prefixes (49.34\%  and 23.61\% out of all ASes and longest matching prefixes present in the BGP routing table, respectively) and 959~666 /24 blocks. 

For the open resolver scan, we retrieve query responses with the \texttt{NOERROR} reply code, meaning that we reached an open resolver. Note that for this study, we do not check the integrity of those responses. In total, we identify  4~022~711 open resolvers, 956~969  of which (23.79\%) are forwarders.

\begin{figure}[t]
  \centering
   \includegraphics[width=0.9\textwidth]{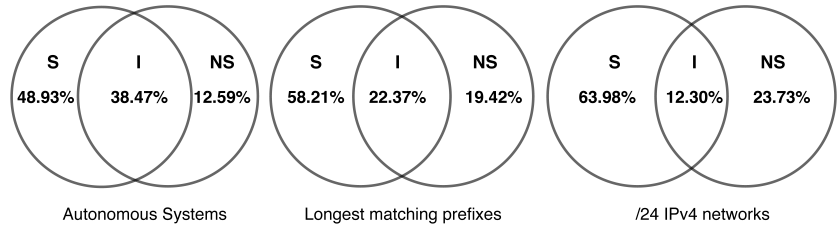}
  \caption{Filtering inconsistencies (S: vulnerable to inbound spoofing, NS: non-vulnerable to inbound spoofing, and I: inconsistent)}
  \label{venn}
\end{figure}

\subsection{Inferring Absence and Presence of Inbound Filtering} 

We compare the results of spoofing and open resolver scans to reveal the absence and the \textit{presence} of inbound SAV. For every detected open resolver, we check whether this particular resolver resolved a spoofed query. If it did not, we assume that this resolver is inside a network performing inbound SAV. We note, however, that due to inconsistent filtering policies inside networks and possible packet losses, we may obtain contradictory results for a single AS or a network. We define ASes and networks as inconsistent if we have at least two measurements with a different outcome. 

Figure~\ref{venn} shows the number of vulnerable to inbound spoofing (S), non-vulnerable to inbound spoofing (NS), and inconsistent (I) ASes, prefixes, and /24 networks. As expected, the most inconsistent results are for ASes with 14~382 (38.47\%) of them revealing both the absence and  presence of inbound SAV. The smaller the network size, the more consistent results we obtain, as it can be seen for the longest matching prefixes and /24 networks. While~/24~is~a common unit of network filtering policy measurement, it still exhibits a high level of inconsistency with 154~704 (12.30\%) networks belonging to both groups.
Importantly, after our initial scan, we ascertained that one of our three upstream providers performed source address validation of outbound traffic. This means that all packets with spoofed source IP addresses routed by this provider were dropped, while those routed by other two upstream providers were forwarded. This has significantly affected the measurements 
and resulted in a very high number of inconsistent results. By disabling this provider before the main scan, the number of inconsistent /24 networks decreased more than two-fold.

\begin{figure}[t]
  \begin{minipage}[b]{0.49\textwidth}
  \includegraphics[width=\textwidth]{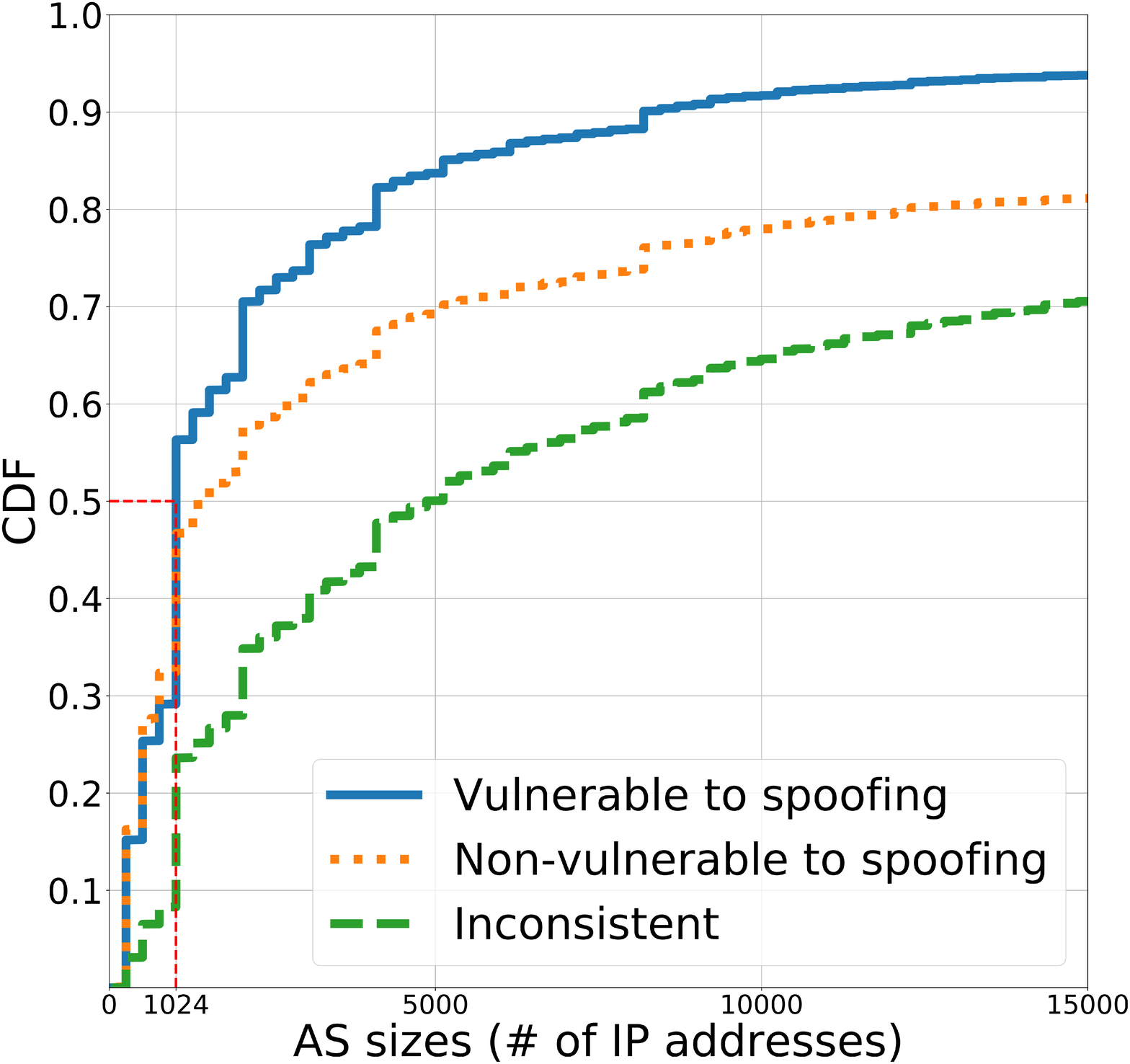}
  \caption{Sizes of autonomous systems}
  \label{cdf_as_sizes}
  \end{minipage}
  \hfill
  \begin{minipage}[b]{0.49\textwidth}
   \includegraphics[width=\textwidth]{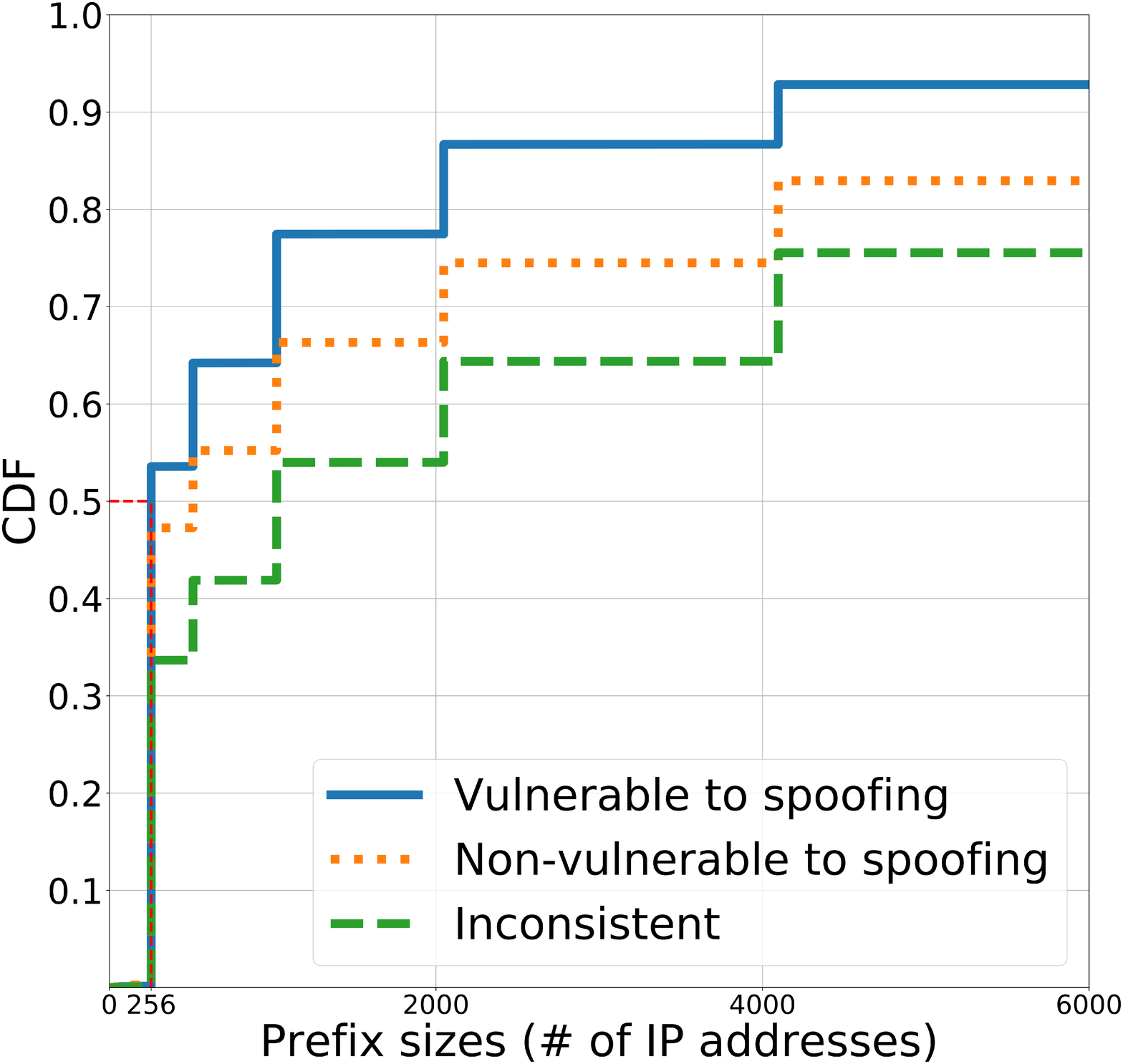}
  \caption{Sizes of longest matching prefixes}
  \label{cdf_prefix_sizes}
  \end{minipage}
  \vspace{-0.2cm}
\end{figure}

Figure~\ref{cdf_as_sizes} presents the cumulative distribution of vulnerable to spoofing, non-vulnerable to spoofing, and
inconsistent AS sizes (the number of announced IPv4 addresses in the BGP routing table). Around 80\% of vulnerable to spoofing ASes have 4~096 addresses and less, meaning that small ASes are less likely to perform packet filtering at the network edge. Figure~\ref{cdf_prefix_sizes} shows the longest matching prefix sizes. We can observe that almost 90\%  of vulnerable to spoofing prefixes have 4~096 addresses and less. The sizes of inconsistent ASes and prefixes are significantly larger compared to vulnerable and non-vulnerable ones.

Larger ASes are more likely to peer with a larger number of other ASes and sub-allocate the address space to their customers and therefore, have less consistent filtering policies. To further understand the complexity of the ASes, we use the CAIDA AS relationship data~\cite{dimitropoulos2007relationships} to compute the number of relationships of all ASes in our dataset. We find that inconsistent ASes maintain relationships with 29 other ASes on average, while vulnerable to spoofing and non-vulnerable to spoofing ASes connect to around 7 ASes on average. 

The AS relationship data and the AS size give us some initial insights into understanding inconsistencies in ASes.
Another possible reason for inconsistent results for a single AS or a network is packet losses. 
To test this hypothesis, we sampled 1000 /24 networks from the inconsistent group and re-scanned them. 
48 networks out of 1000 did not respond to 
any query. Most importantly, 464 became consistent (all vulnerable to spoofing). The remaining /24s were still inconsistent.

Furthermore, we repeated the same test once per day, three days in a row, to estimate the persistence of the results. More than two-thirds of the networks belonged to the same group (inconsistent, vulnerable to spoofing, non-vulnerable, or no data) during all three measurements. Interestingly, half of those were inconsistent. For most of the networks, the exact set of the responding resolver IP addresses changed every day, 
due to the IP address churn of resolvers~\cite{Kuhrer:wild}. Regarding the remaining one-thirds, it is unlikely that provider filtering policies change so rapidly. Therefore, apart from packet losses, we may be dealing with other issues such as different filtering policies at upstream providers for multi-homed customer ASes.

These experiments show that even though the number of inconsistent /24s decreased 
almost two-fold, such networks are not uncommon. We plan to contact several network providers to validate our results and gain some insights into their motivation for inconsistent filtering at the network level.


\subsection{Geographic Distribution}

Identifying the countries that do not comply with the SAV standard is the first step in mitigating the issue by, for example, contacting local CSIRTs.
We use the MaxMind database\footnote{https://dev.maxmind.com/geoip/geoip2/geolite2/} to map every resolver IP address of the spoofed query retrieved from the domain name to its country. Table~\ref{geolocation} summarizes the results.  

\begin{table}[t]
\caption{Geolocation results} 
\label{geolocation} 
\scriptsize
\centering 
\setlength{\tabcolsep}{2pt}
\renewcommand{\arraystretch}{1}
\setlength{\arrayrulewidth}{0.5pt}
\begin{tabular}{c|lc|lc|lc}
\thickhline
 &&&&\textbf{Vulnerable to}&&\textbf{Vulnerable to}\\
\textbf{Rank}&\textbf{Country}&\textbf{Resolvers (\mypound)}&\textbf{Country}&\textbf{spoofing /24}&\textbf{Country}&\textbf{spoofing /24}\\
 &&&&\textbf{networks (\mypound)}&&\textbf{networks (\%)}\\
\thickhline
 1 & China & 2~304~601 & China & 271~160 & Cocos Islands & 100.0 \\
 2 & Brazil & 687~564 & USA & 157~978 & Kosovo & 81.82 \\
 3 & USA & 678~837 & Russia & 55~107 & Comoros & 57.89 \\
 4 & Iran & 373~548 & Italy & 32~970 & Armenia & 52.16 \\
 5 & India & 348~076 & Brazil & 29~603 & Western Sahara & 50.00 \\
 6 & Algeria & 252~794 & Japan & 28~660 & Christmas Island & 50.00 \\
 7 & Indonesia & 249~968 & India & 27~705 & Maldives & 39.13 \\
 8 & Russia & 229~475 & Mexico & 24~195 & Moldova & 38.66 \\
 9 & Italy & 108~806 & UK & 18~576 & Morocco & 37.85 \\
 10 & Argentina & 103~449 & Morocco & 18~135 & Uzbekistan & 36.17 \\
\thickhline
\end{tabular}
\end{table}

In total, we identified 237 countries and territories vulnerable to spoofing of incoming network traffic. We first compute the number of DNS resolvers per country and map the resolvers to the nearest /24 IP address blocks to evaluate the number of vulnerable to spoofing /24 networks per country. We see that the top 10 countries by the number of DNS resolvers are not the same as the top 10 for vulnerable to spoofing /24 networks because a large number of individual DNS resolvers by itself does not indicate how they are distributed across different~networks.

Such absolute numbers are still not representative as countries with a large Internet infrastructure may have many DNS resolvers and therefore reveal many vulnerable to spoofing networks that represent a small proportion of the whole. For this reason, we compute the fraction of vulnerable to spoofing vs. all /24 IPv4 networks per country. To determine the number of all the /24 networks per country, we map all the individual IPv4 addresses to their location, then to the nearest /24 block and keep the country/territory where most addresses of a given network belong to. Figure~\ref{spoofers_map} (in Appendix) presents the resulting world map. We can see in Table \ref{geolocation} that the top 10 ranking has changed. Small countries such as Cocos Islands and Western Sahara, which have one identified resolver each, suffer from a high proportion of vulnerable to spoofing networks. The only /24 network of Cocos Islands allows inbound spoofing. On the other hand, Morocco is a country with a large Internet infrastructure (47~915 /24 networks in total) and with a large relative number of vulnerable to spoofing networks.


\subsection{Outbound vs. Inbound SAV Policies \label{sec:policies}}
We retrieve the Spoofer data to infer the absence and the presence of outbound SAV. The Spoofer client sends packets with the IP address of the machine on which the client is running as well as packets with a spoofed source address. The results are anonymized per /24 IP address blocks. Spoofer identifies four possible states: \textit{blocked} (only an unspoofed packet was received, the spoofed packet was blocked), \textit{rewritten} (the spoofed packet was received, but its source IP address was changed on the way), \textit{unknown} (neither packet was received), \textit{received} (the spoofed packet was received by the~server).

In December 2019, we collected and aggregated the results of the latest measurements  to infer outbound SAV compliance. During this period, tests were run from 3~251 distinct /24 networks. In some cases, tests from the same IP address block show different results due to possible changes in the filtering policies of the tested networks, so we kept the latest result for every /24 network. We identified 1~910 networks blocking spoofed outbound traffic and 316 that allow spoofing.

We deploy the forwarders-based technique on our scanning server and analyze the responses in which the originally queried IP address is not the same as the responding one, as described in Section \ref{sec:orscan}.
Interestingly, 3~147 responses arrived from the private ranges of IP addresses. Previous work has shown that this behavior is related to NAT misconfiguration~\cite{spoofer_new}. 
To detect misbehaving forwarders inside networks vulnerable to outbound spoofing, we check that the IP address of the forwarder, the source IP address of the response, and the scanner IP address belong to different ASes.
In this way, we identify  456~816 misbehaving forwarders originated
from 20~889 /24 IP networks vulnerable to outbound spoofing. In total, the two methods identify 21~203 /24 networks without outbound filtering and 1~910 /24 networks with outbound SAV in place.

The results obtained by running our scans and using the data of the Spoofer project let us evaluate the filtering policies of networks in both directions (inbound and outbound). We aggregated all the datasets and found 4~541 /24 networks with no filtering in any direction and only 151 networks deploying both inbound and outbound SAV. To further understand the filtering preferences of network operators, 
we analyze how many of them do not filter only outbound or only inbound traffic. Note however, that the coverage of inbound filtering scans is much larger than the one of outbound SAV (forwarders-based method and especially the Spoofer data). To make the datasets comparable, we find the intersection between the networks covered by inbound filtering scans (only those showing consistent results) and all the networks tested with the Spoofer client. In the resulting set of 559 /24 networks, there are 298 networks with no filtering for inbound traffic only and 15 with no outbound SAV only. It shows that inbound filtering is less deployed than outbound, which is consistent with previous work \cite{spoofer_new}. We do the same comparison of our inbound filtering scans and the forwarders-based method. Among 16~839 common /24 networks (all vulnerable to outbound spoofing), 12~393 are also vulnerable to inbound spoofing.

\section{Limitations}

While we aimed at designing a universal method to detect the deployment of inbound SAV at the network edge, our approach has certain limitations that may impact the accuracy of the obtained results. We rely on one main assumption---the presence of an (open or closed) DNS resolver  or a proxy in a tested network. In case of the absence of one of them, we cannot conclude on the filtering policies. If the probed resolver is closed, our method may only confirm that a particular network does not perform SAV for inbound traffic, at least for some part of its IP address space. Only the presence of an open DNS resolver may reveal the SAV presence assuming that the transit networks do not deploy SAV.

From our results, we often cannot unequivocally conclude on the general policies of operators of, for example, larger autonomous systems. Some parts of an AS, a BGP prefix, or even a /24 network may be configured to allow spoofed packets to enter one subnetwork and to filter spoofed packets in another one.

The scanner sending spoofed packets should itself be located in the network not performing SAV for outgoing traffic. Still, even if a spoofed query leaves our network, it may be filtered by some transit networks and never reach the tested destination. Therefore, we plan to test our method from different vantage points.

There are several reasons, apart from deploying SAV, why we have no data for certain IP address blocks. Packet losses and temporary network failures are some of the reasons for not receiving queries from all the target hosts~\cite{Kuhrer:wild}. To overcome this limitation, we plan to repeat our measurements regularly and study the persistence of this vulnerability over time.

\section{Conclusion}

The absence of ingress filtering at the network edge may lead to important security issues such as DDoS and DNS zone or cache poisoning attacks. Even if network operators are aware of these risks, they choose not to filter traffic, or do it incorrectly because of  deployment and maintenance costs or implementation issues. There is a need for identifying and enumerating networks that do not comply with RFC 2827 to understand the scale of the problem and find possible ways to mitigate it. 
 
In this paper, we have presented a novel method for inferring the SAV deployment for inbound traffic and discussed the results of the first Internet-wide measurement study. We have obtained significantly more test results than other methods: we cover over 49\% of autonomous systems and 23\% of the longest matching BGP prefixes.

The next step for this work is to start longitudinal measurements to infer the SAV deployment in both IPv4 and IPv6 address spaces from different vantage points. Finally, we plan to notify all parties affected by the vulnerability.
\newpage

\noindent \textbf{Acknowledgments.}
The authors would like to thank the anonymous
reviewers and our shepherd Ramakrishna Padmanabhan for their valuable feedback.
This work has been carried out in the framework of the PrevDDoS project funded by
the IDEX Universit\'e Grenoble Alpes ``Initiative de Recherche Scientifique (IRS)''. 

\section*{Appendix}

\begin{figure}
\centering
\includegraphics[width=0.9\textwidth]{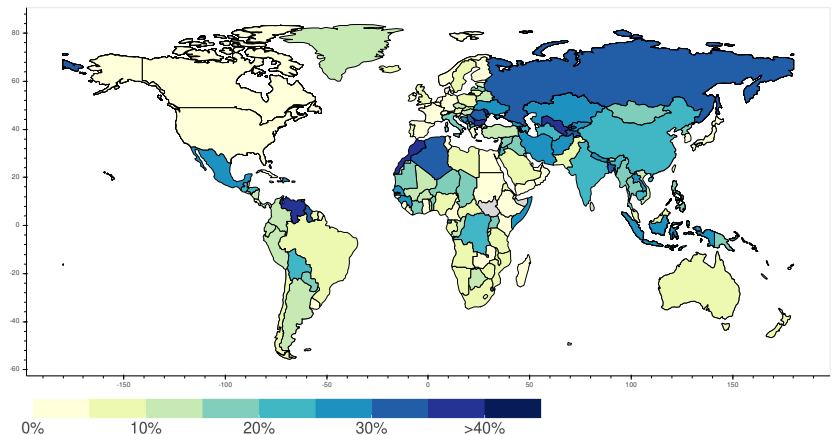}
\caption{Fraction of vulnerable to spoofing (inbound traffic) vs. all /24 networks per country}
\label{spoofers_map}
\vspace{-0.2cm}
\end{figure}

\bibliographystyle{splncs04}
\bibliography{references}

\end{document}